\documentclass[conference]{IEEEtran}

\makeatletter
\def\endthebibliography{%
	\def\@noitemerr{\@latex@warning{Empty `thebibliography' environment}}%
	\endlist
}
\makeatother

\usepackage{ifpdf}

\usepackage{cite}
\usepackage{subcaption}
\usepackage[abs]{overpic}

\usepackage{graphicx}
\usepackage{epstopdf}

\usepackage[cmex10]{amsmath}
\usepackage{amsfonts,amssymb}
\usepackage{xcolor}

\usepackage{algpseudocode}
\usepackage{algorithm}

\usepackage{array}

\usepackage{expl3}

\usepackage{dblfloatfix}

\usepackage{newtxtext, newtxmath}
\usepackage{url}

\usepackage{arydshln}
\usepackage{enumerate}
\usepackage[mathcal]{euscript}

\newtheorem{lemma}{Lemma}

\newtheorem{assumption}{\textbf{Assumption}}
\newtheorem{remark}{\textbf{Remark}}
\newcommand{\vect}[1]{\mathbf{#1}}

\def\beq{\begin{equation}}
\def\eeq{\end{equation}}

\graphicspath {{Figures/}}
\IEEEoverridecommandlockouts

\begin{document}
%
\title{NOMA Versus Massive MIMO in Rayleigh Fading}

\author{\IEEEauthorblockN{Kamil Senel, Hei Victor Cheng, Emil Bj\"{o}rnson, and Erik G.~Larsson} 
	\IEEEauthorblockA{Dept. of Electrical Engineering, Link\"{o}ping University, Link\"{o}ping, Sweden\\
		Email: \{kamil.senel, emil.bjornson, erik.g.larsson\}@liu.se}
	\thanks{This work was supported in part by ELLIIT and the Swedish Research Council (VR).}
}

\maketitle


\begin{abstract}
This paper compares the sum rates and rate regions achieved by power-domain NOMA (non-orthogonal multiple access) and standard massive MIMO (multiple-input multiple-output) techniques. We prove analytically that massive MIMO always outperforms NOMA in i.i.d.~Rayleigh fading channels, if a sufficient number of antennas are used at the base stations. The simulation results show that the crossing point occurs already when having 20-30 antennas, which is far less than what is considered for the next generation cellular networks.
\end{abstract}


\IEEEpeerreviewmaketitle

\section{Introduction}\label{sec:Introduction}

Two main technologies have been considered to improve the spectral efficiency of 5G networks: non-orthogonal multiple access (NOMA) \cite{ding2017survey} and massive multiple-input multiple-output (mMIMO) \cite{redbook}. Do we need to choose between them or can their respective gains be combined?

The key idea of (power-domain) NOMA is to serve multiple users at the same time/frequency/code resource, typically by dividing the users into groups and superimposing the data signals within each group. When multiple antennas are used, the same beamforming is used within a group.
 In each group, the user with the best channel can decode the signal sent to the users with the poorer channels, and successive interference cancellation (SIC) is used to eliminate the interference \cite{ding2017application}. 

mMIMO refers to systems where the base stations (BSs) are equipped with a large number of antennas, $M$, as compared to the number of users, $K$. 
By having $M \gg K$, the BS has the spatial resolution to spatially multiplex the users on the same time/frequency/code resource \cite{redbook,massivemimobook}. 
Each user is assigned an individual beam, based on its channel, and spatial interference-suppression such as zero-forcing (ZF) beamforming is used to limit inter-user interference. mMIMO is already used in LTE-Advanced and will be a core feature of 5G.

While the vast majority of prior works on NOMA considers single-antenna BSs, there are a few papers with $M$-antenna BSs. 
A comparison of NOMA and multi-user transmission can be found in \cite{chen2016application,chen2016beamforming} for $M \approx K$, which is not a mMIMO setup, and in \cite{cheng2018performance} for $M \gg K$ but using the vastly suboptimal maximum-ratio processing scheme. To the best of our knowledge, the performance of NOMA has not been compared with mMIMO in setups with $M \gg K$, state-of-the-art mMIMO methods (such as ZF), and imperfect channel knowledge.

\textbf{\textit{Main contributions:}} In this paper, we compare NOMA and mMIMO in a single-cell setup with arbitrary $M$ and $K$, assuming i.i.d.~Rayleigh fading channels and ZF. 
The analysis shows that mMIMO outperforms NOMA in terms of sum rate.
We derive closed-form expressions for the maximum sum rate in a two-user setup and obtain the minimum number of BS antennas required for mMIMO  to outperform NOMA.

\section{System Setup}\label{sec:SystemSetup}

We consider the downlink transmission of a single-cell system with an $M$-antenna BS and $K$ single-antenna users. We consider block fading channels, where the time-frequency resources are divided into coherence intervals in which the channels are constant and frequency flat. 
We let $T$ denote the number of modulation symbols per coherence interval.
The system operates in time-division duplex (TDD) mode so that channel reciprocity can be utilized at the BS to estimate the downlink channels based on uplink pilots, and the BS later uses these estimates for downlink multiuser beamforming.

The channel vector $\vect{g}_k = \sqrt{\beta_k} \vect{h}_k  \in \mathbb{C}^{M}$ of user $k$ has the large-scale fading coefficient $\beta_k\geq 0$ and the small-scale fading vector $\vect{h}_k \in \mathbb{C}^{M}$. The latter takes one independent i.i.d.~Rayleigh fading realization in each coherence interval:
\begin{equation}\label{eq:RayleighFading}
\vect{h}_k \sim \mathsf{CN}\,(\vect{0}, \vect{I}_M), \quad k = 1, \ldots,K.
\end{equation} 
The user set, $\mathcal{K}$, consists of $K/2$ cell-edge users and $K/2$ cell-center users, where $K$ is even. The indices $\mathcal{K}_c = \{1, \ldots, K/2 \}$ are utilized for cell-center users and $\mathcal{K}_e = \{K/2+1, \ldots, K\}$ denotes cell-edge users. This classification is made such that  
\begin{equation}\label{eq:largeScaleAssumption}
\beta_j > \beta_i, \quad \forall j \in \mathcal{K}_c,~~ \forall{i} \in \mathcal{K}_e.
\end{equation}
We will compare the ergodic rates achieved by a typical NOMA scheme and a typical mMIMO scheme. Since the user grouping must be the same for many channel realizations to enable decoding, it only depends on the  large-scale fading.

\vspace{-1mm}

\subsection{NOMA Scheme} 

\vspace{-1mm}

We consider a \emph{NOMA scheme} that has been analyzed in \cite{cheng2018performance,kim2013non}. The $K$ users are divided into $K/2$ groups, where each group consists of one cell-center and one cell-edge user. Without loss of generality, we assume that the indices $i$ and $i + K/2$ denote the users in group $i$ for $i = 1, \ldots, K/2$.
The interference between the groups is suppressed by beamforming. More precisely, ZF beamforming vectors are selected to cancel interference between the cell-center users \cite{chen2017exploiting,kim2013non}, i.e.,
\begin{equation}\label{eq:conditionBF}
\vect{v}_i^H\vect{h}_j = 0,\hfill~~~\text{if}~j\neq i, \quad \forall i, j = 1,\ldots,K/2,
\end{equation}
 where $\vect{v}_i$ is the beamforming vector for group $i$.
 The rationale for this choice is that the cell-center users are sensitive to interference from other groups since they need to perform SIC to cancel interference from the cell-edge user in their own group. To implement this beamforming, the BS only needs to know the $K/2$ channels to the cell-center users. 
 
\begin{figure} 
	\begin{tabular}{cc}
		\begin{subfigure}{.2\textwidth}
		\includegraphics[width=40mm]{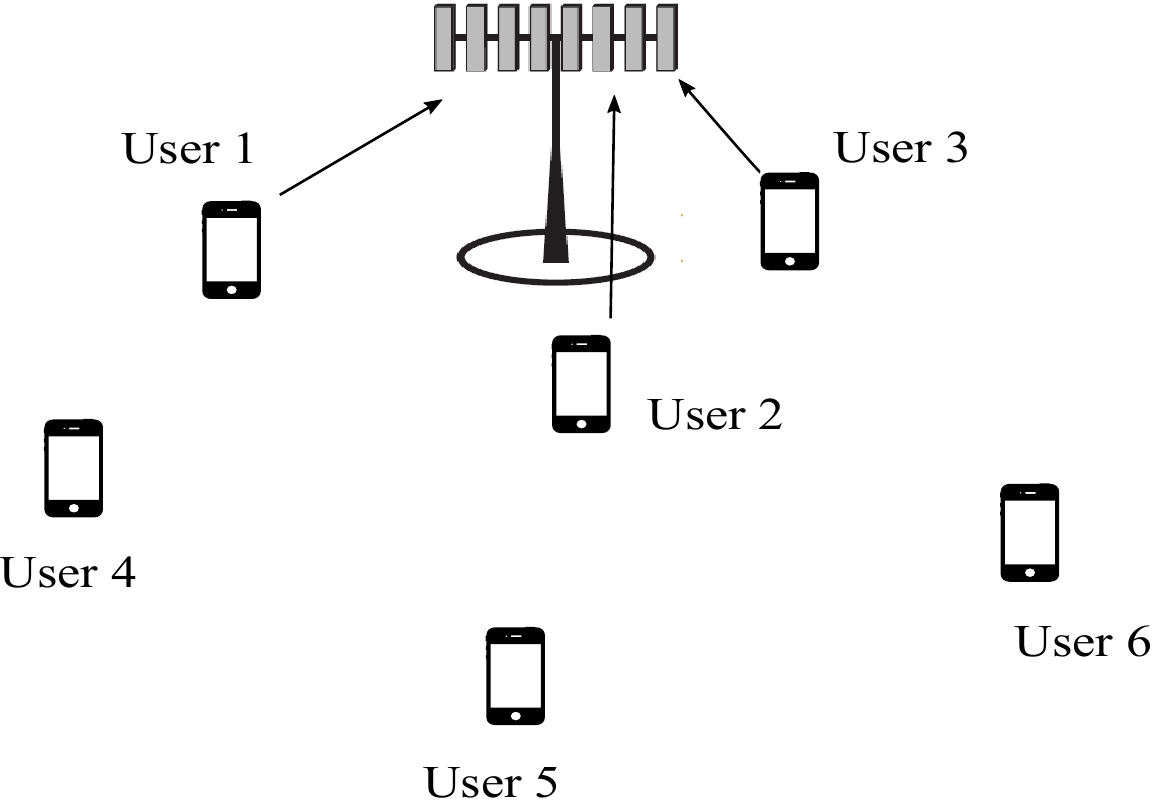}
		\caption{} \label{fig:NOMAtraining}
	\end{subfigure} & \hfill \begin{subfigure}{.2\textwidth}\begin{center} 
	\includegraphics[width=40mm]{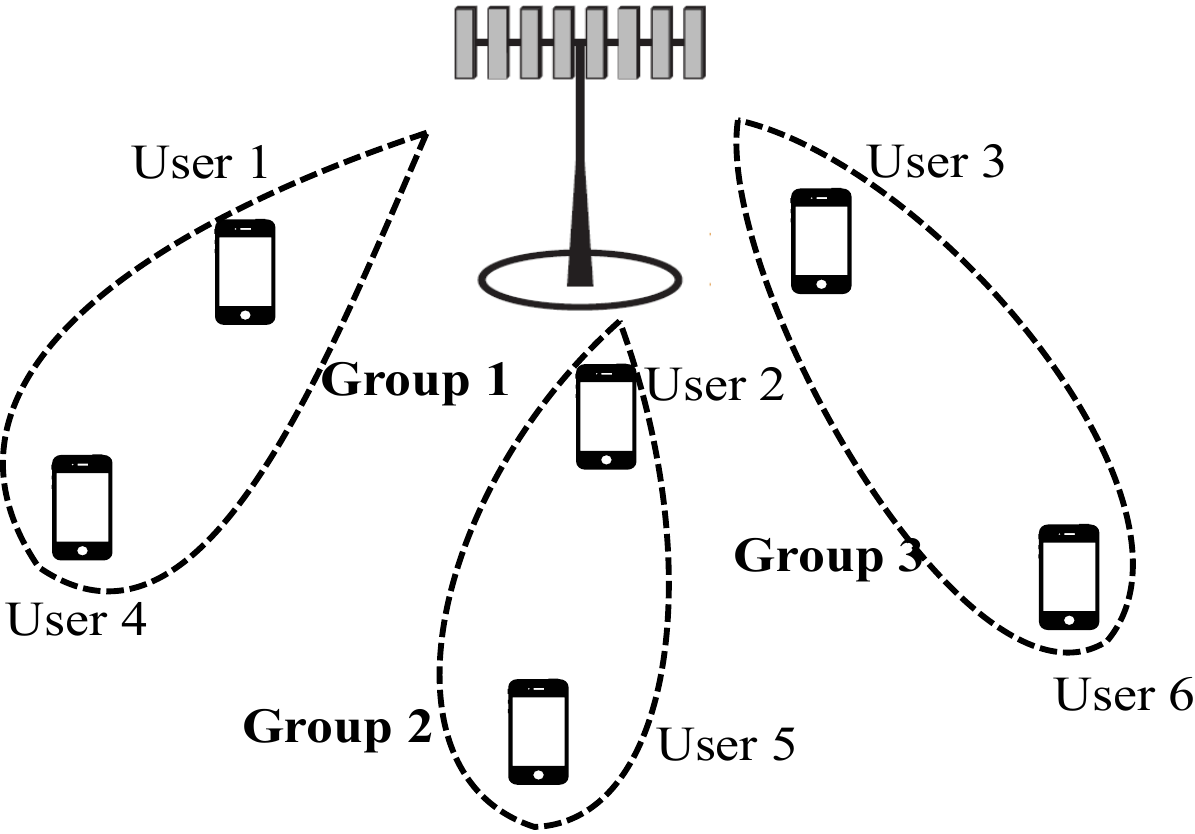}
	\caption{} \label{fig:NOMAdata} \end{center}
\end{subfigure} \\ \vspace{-1mm} & \vspace{-1mm}\\
		\begin{subfigure}{.2\textwidth}
			\includegraphics[width=40mm]{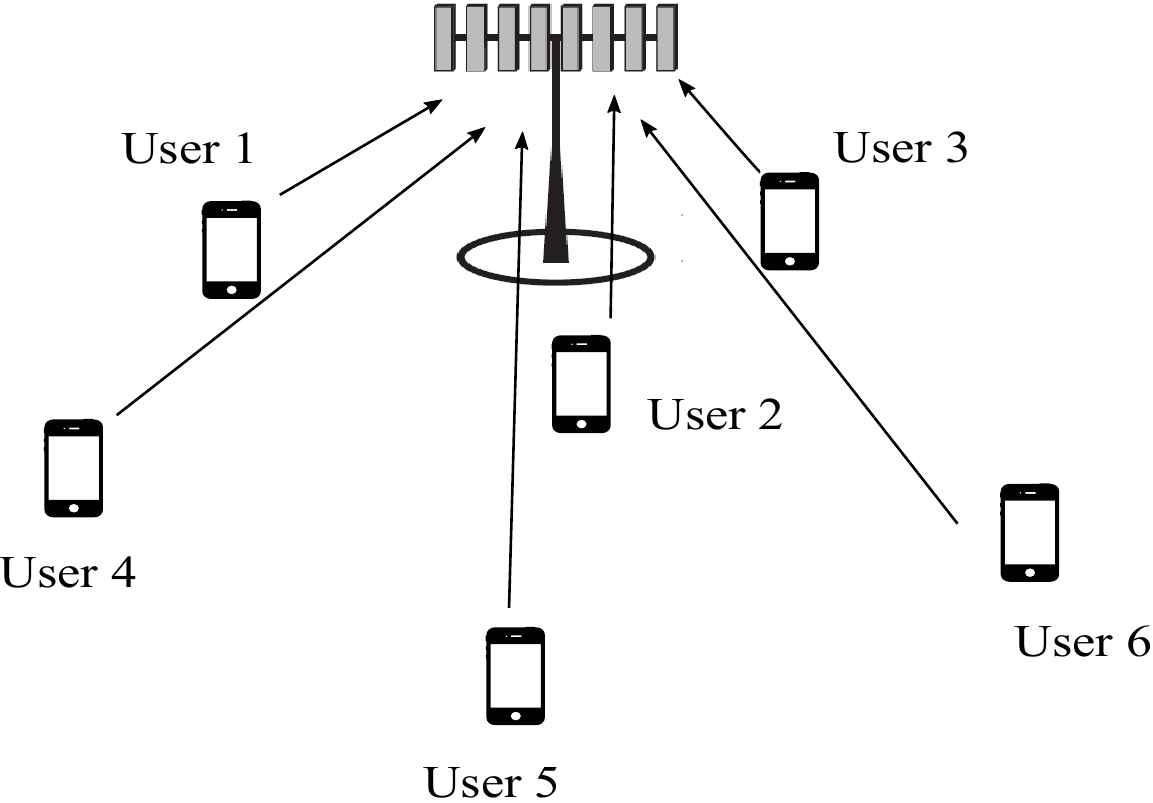}
			\caption{} \label{fig:ZFtraining}
		\end{subfigure} & \hfill \begin{subfigure}{.2\textwidth}
		\includegraphics[width=40mm]{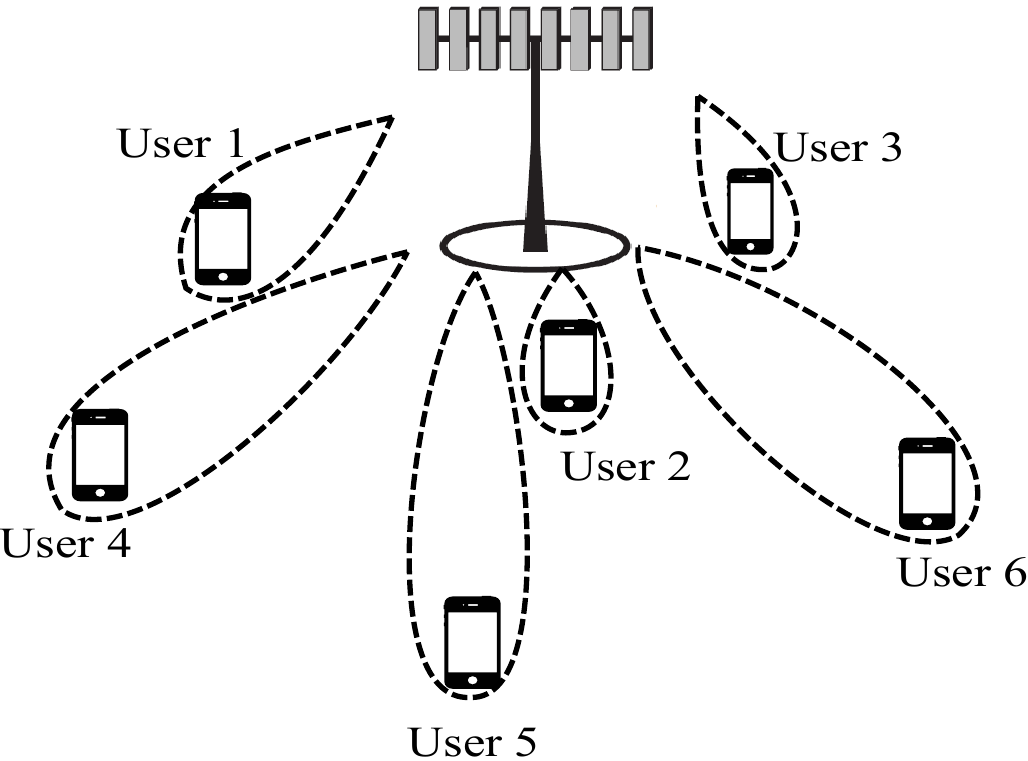}
		\caption{} \label{fig:ZFdata}
	\end{subfigure} 
	\end{tabular}
	\caption{The training and data transmission phases for mMIMO and NOMA. (a) During the training of NOMA cell-center users transmit pilot sequences. (b) Users in the same NOMA group receives data via the same beamforming vector. (c) In mMIMO, each user transmits its pilot sequence during training. (d) Each user has a separate mMIMO beamforming vector generated by using the estimate of their channels.   } \vspace{-3mm}
	\label{fig:NOMA-ZFsetups}
\end{figure}

\subsection{mMIMO Scheme} 

The considered \emph{mMIMO scheme} is also based on ZF beamforming, which is known to perform well in single-cell mMIMO systems with many antennas \cite{redbook}. In the mMIMO case, unlike NOMA, there are no groups. There are $K$ beamforming vectors, where $\vect{v}_k$ is the beamforming associated with user $k$, and these are selected based  on all the $K$ channels instead of $K/2$ channels, as in NOMA. Hence, the BS needs to know these $K$ channels. 
Although $M\gg K$ is typical for mMIMO systems, the analysis in this paper hold for any $M,K$.

\begin{remark} \label{remark-NOMA}
Throughout the paper, we will make a series of assumptions that are beneficial to NOMA as our goal is to find conditions where NOMA might perform better than the mMIMO scheme. Since the results are biased in favor of NOMA, we can be sure that mMIMO outperforms NOMA whenever the analytical results show that. 
\end{remark}

\subsection{Pilot Overhead for Channel Estimation}\label{sec:ChEstimation}

Each coherence interval consists of two phases: uplink/downlink training and downlink data transmission. During the uplink training, some or all of the users transmit pilot sequence and the BS estimates the channels. These estimates are used to generate the beamforming during data transmission. 

The NOMA scheme requires $K/2$-length pilot sequences in the uplink: one orthogonal pilot per cell-center user. The training and data transmission phases are illustrated in Fig.~\ref{fig:NOMA-ZFsetups}(a)-(b) in an example with $K = 6$ users. 
Since the beamforming vectors are only based on the cell-center users' channels, the effective channels of the cell-edge users can have any phase and magnitude. Hence, the BS needs to send $K/2$ downlink pilot sequences to enable downlink channel estimation and coherent detection. We make the following simplifying assumption.

\begin{assumption}\label{as:DLpilots}
In the analysis of the NOMA scheme, perfect CSI is acquired at the users via $K/2$ downlink pilots. 	
\end{assumption}

The mMIMO scheme requires $K$-length pilot sequences in the uplink: one orthogonal pilot per user. The training and data transmission phases are illustrated in Fig.~\ref{fig:NOMA-ZFsetups}(c)-(d). 
An important advantage of mMIMO  is that downlink pilots are not needed since the effective channels that are created by the beamforming are highly predictable, as explained in \cite{ngo2017no, massivemimobook}. We assume minimum-mean square error (MMSE) estimation is used at the BS.

In summary, $K$ pilots are needed for both schemes, but they are allocated differently between uplink and downlink.

\section{Performance Analysis} \label{sec:NLOS}

In this section, we analyze and compare the achievable rates of the considered NOMA and mMIMO schemes. 
\subsection{Downlink Data Transmission}

After the training phase, the BS generates beamforming vectors, as described in Section~\ref{sec:SystemSetup},
and user $k$ receives
\begin{equation}\label{eq:RecSignal}
y_k = \sum_{k' = 1}^{K} \vect{g}_k^T\vect{x}_{k'} + z_k = \sum_{k' = 1}^{K} \sqrt{\beta_k} \vect{h}_k^T\vect{x}_{k'} + z_k,
\end{equation}
where $z_k \sim \mathsf{CN}\,(0,\,1)$ is the additive noise and $\vect{x}_k \in \mathbb{C}^M$ is the beamformed data symbol of user $k$, obtained as
\begin{equation}
\vect{x}_k = \vect{v}_k \sqrt{p_k}s_k,
\end{equation}
where $s_k\sim\mathsf{CN}(0, 1)$ is the data symbol of user $k$, $p_k$ is the normalized transmission power, and $\vect{v}_k$ is the beamforming vector (satisfying $\mathbb{E}\{\|\vect{v}_k\|^2\} = 1$). This basic model applies to both the NOMA and mMIMO schemes, but the beamforming vectors are selected differently.
In the NOMA scheme, the beamforming vector of the users in the same group are identical, i.e., $\vect{v}_k = \vect{v}_{k + K/2}$ for all $k = 1, \ldots, K/2$. In contrast, each user has a unique beamforming vector in the mMIMO scheme. 

\subsubsection{Rates with NOMA}

Since the users are assumed to have perfect CSI, the instantaneous SINR of user $k$ is 
\begin{equation} \label{eq:SINRNOMAnoSIC}
\text{SINR}_k = \frac{p_k\beta_k|\vect{h}_k^T \vect{v}_{k}|^2}{\beta_k \sum_{k' \neq k}^{K} p_{k'}|\vect{h}_k^T \vect{v}_{k'}|^2 + 1}.
\end{equation} 

Each cell-edge user treats the interference as noise and decodes only its own data symbols, whereas each cell-center user decodes the data symbols of the cell-edge user in the same group and performs SIC, hence, effectively removing the interference from the cell-edge user. However, in order to perform SIC, the cell-center user needs to be able to decode data signal intended for cell-edge user, i.e., the ergodic achievable rate of the data signal of the cell-edge user, $s_{k+K/2}$, at user $k$ must be greater than or equal to the ergodic achievable rate of the corresponding cell-edge user: 
\begin{equation}\label{eq:SICcond}
\mathbb{E}\left[\log_2 \left(1 + \text{SINR}_{k,k + K/2} \right)\right]  \geq \mathbb{E}\left[\log_2 \left(1 + \text{SINR}_{k + K/2} \right)\right],
\end{equation} 
where 
\begin{equation}
\text{SINR}_{k,k + K/2} =  \frac{p_{k+K/2}\beta_{k}|\vect{h}_k^T \vect{v}_{k + K/2}|^2}{\beta_k \sum_{k' \neq k + K/2}^{K} p_{k'}|\vect{h}_k^T \vect{v}_{k'}|^2 + 1}.
\end{equation}
This condition can always be satisfied by selecting the transmit powers appropriately, thus we make the following assumption.

\begin{assumption}\label{as:SICcond}
	In the NOMA scheme, the SIC condition in \eqref{eq:SICcond} is satisfied for each group. 
\end{assumption}

Under Assumptions \ref{as:DLpilots} and \ref{as:SICcond}, the achievable ergodic rate of cell-center user $k$ is given by \cite{cheng2018performance} \vspace{-1mm}
\begin{flalign}
R_{k}^{\mathrm{NOMA}}=\tau  \mathbb{E}\left[\log_2\left(1+\frac{p_k
	\beta_k |\vect{h}_k^T\vect{v}_k|^2}{\beta_k\hspace{-3mm}\sum\limits_{\substack{k'\neq k,\\ k' \neq k+K/2}}^K \hspace{-3mm} p_{k'}|\vect{h}_k^T\vect{v}_{k'}|^2+ 1}\right)\right], \forall k \in \mathcal{K}_c,\label{eq:rateNOMAcc}
\end{flalign} \vspace{-3mm}

\noindent where $\tau = \left(1-\frac{K}{T}\right)$ is the fraction of each coherence interval  used for data.

For the cell-edge users, the achievable rate is given by 
\begin{equation}\label{eq:rateNOMAce}
R_{k}^{\mathrm{NOMA}}=\tau\mathbb{E}\left[\log_2 \left(1 + \text{SINR}_{k + K/2} \right)\right], \forall k \in \mathcal{K}_e,
\end{equation} 
where $\text{SINR}_{k + K/2}$ is defined in \eqref{eq:SINRNOMAnoSIC} \cite{cheng2018performance}. 
These rate expressions must be evaluated numerically, but we can find closed-form upper bounds as follows.

\begin{assumption}\label{as:InterferenceGroups}
	For the NOMA scheme, the beamforming is assumed to perfectly eliminate interference between groups. 
\end{assumption}

Using Assumption \ref{as:InterferenceGroups}, we obtain the upper bounds 
\begin{flalign}
R_{k}^{\mathrm{NOMA}} &\leq\tau  \mathbb{E}\left[\log_2\left(1+p_k
	\beta_k |\vect{h}_k^T\vect{v}_k|^2\right)\right] \label{eq:rateNOMAccUpperBound} \\
	&\leq\tau  \log_2\left(1+p_k
	\beta_k \mathbb{E}\left[|\vect{h}_k^T\vect{v}_k|^2\right]\right) , \forall k \in \mathcal{K},_c\label{eq:rateNOMAccUpperBound-2}
\end{flalign}
where \eqref{eq:rateNOMAccUpperBound-2} is due to Jensen's inequality, and similarly
\begin{flalign}R_{k}^{\mathrm{NOMA}} &\leq\tau \mathbb{E}\left[\log_2\left(1+\frac{p_k
	\beta_k |\vect{h}_k^T\vect{v}_k|^2}{\beta_k p_{k-K/2}|\vect{h}_k^T\vect{v}_k|^2 + 1}\right)\right], \label{eq:rateNOMAceUpperBound} \\
	& \leq\tau  \log_2\left(1+\frac{p_k
	\beta_k \mathbb{E}\left[|\vect{h}_k^T\vect{v}_k|^2\right]}{\beta_k p_{k-K/2}\mathbb{E}\left[|\vect{h}_k^T\vect{v}_k|^2\right] + 1}\right), \forall k \in \mathcal{K}_e. \label{eq:rateNOMAceUpperBound-2}
\end{flalign}

\subsubsection{Rates with mMIMO}

The mMIMO scheme has been throughly investigated in the literature \cite{redbook}. 
For any $M\geq K$, an exact ergodic achievable rate for user $k$ is given by
\begin{equation}\label{eq:rateZF}
R_{k}^{\mathrm{mMIMO}} \geq \tau\log_2\left(1 + \frac{\left(M-K\right)p_k\beta_{k}\gamma_{k}^{\mathrm{mMIMO}}}{ \beta_{k}(1 - \gamma_k^{\mathrm{mMIMO}})\sum\limits_{k'= 1}^K p_{k'}+1}\right),
\end{equation}
where the normalized variance of the channel estimate is $\gamma_{k}^{\textrm{mMIMO}} =  \frac{K\beta_kq_k}{K\beta_{k}q_{k} + 1}  \in [0,1]$, with $q_k$ being the uplink pilot power. The parameter $\tau$ is the same as for NOMA, but obtained differently as explained in Section~\ref{sec:ChEstimation}.

\subsection{Performance Comparison: Perfect CSI}

In this subsection, the BS is assumed to obtain the channel estimates perfectly, i.e., $\gamma_{k}^{\mathrm{mMIMO}} = 1$ and \eqref{eq:rateZF} becomes \cite{redbook}
\begin{equation}\label{eq:rateZFperfectCSI}
R_{k}^{\mathrm{mMIMO}} \geq \tau\log_2\left(1 + \left(M-K\right)p_k\beta_{k}\right),
\end{equation}
whereas \eqref{eq:rateNOMAccUpperBound-2} and \eqref{eq:rateNOMAceUpperBound-2} will respectively become
\begin{flalign}
&R_{k}^{\mathrm{NOMA}}\leq\tau  \log_2\left(1+p_k	\beta_k \left(M+1-K/2\right)\right), \forall k \in \mathcal{K}_c, \label{eq:rateNOMAccUpperBound-3}
\end{flalign}
and
\begin{flalign}
&R_{k}^{\mathrm{NOMA}}\leq\tau  \log_2\left(1+\frac{p_k
	\beta_k }{p_{k-K/2}\beta_k + 1}\right), \forall k \in \mathcal{K}_e.\label{eq:rateNOMAceUpperBound-3}
\end{flalign}

The mMIMO scheme eliminates interference completely at each user. On the other hand, NOMA eliminates interference at only one of the users in each group, namely the cell-center user. Moreover, all the $K$ users have a rate that grows with $M$ (known as the array gain) in the mMIMO scheme, while that only happens for the cell-center users in the NOMA scheme.

	\begin{figure}[tb]
	\begin{center} \vspace{-2mm}
		\includegraphics[trim=0.5cm 0cm 0cm 0cm,clip=false, scale = .6]{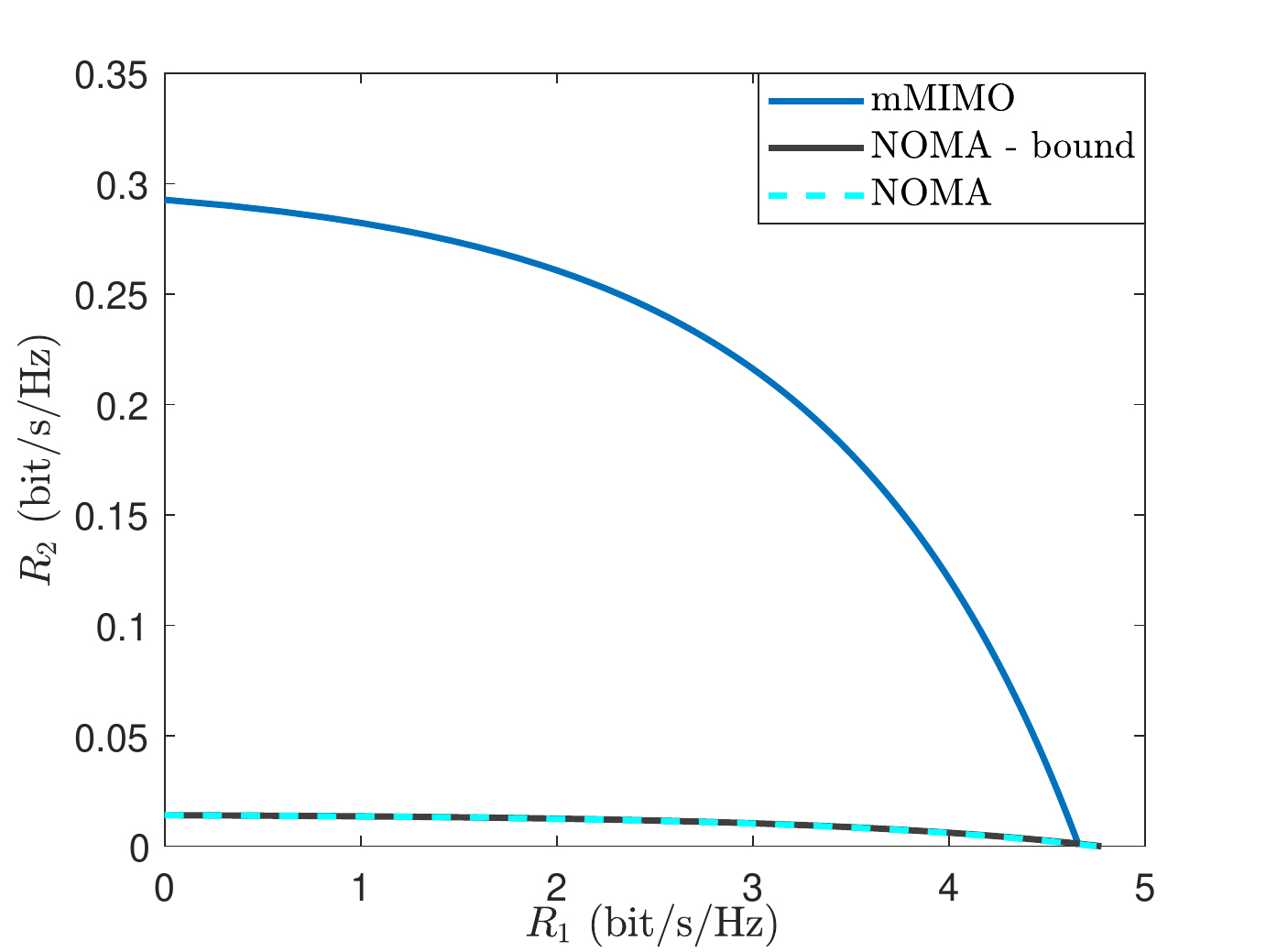}
		\caption{Rate regions obtained for two users with $M = 25$ BS antennas. The curves are obtained for a coherence interval length $T = 100$ symbols under the perfect CSI assumption. }
		\label{fig:Fig1} 
	\end{center} \vspace{-5mm}
\end{figure}

Fig.~\ref{fig:Fig1} illustrates the rate regions for the mMIMO and NOMA schemes for $K = 2$ users and $M = 25$ BS antennas. When the users are assigned full power, the 
cell-center user has a signal-to-noise ratio (SNR) of $15$\,dB and the cell-edge user has $-5$\,dB.
In the figure, the rate regions are obtained using the closed-form rate bounds given in \eqref{eq:rateZF}, \eqref{eq:rateNOMAccUpperBound-3}, and \eqref{eq:rateNOMAceUpperBound-3}, and for NOMA they are compared with the actual rates given by \eqref{eq:rateNOMAcc}, and \eqref{eq:rateNOMAce}. In the rest of the analysis, the lower bound on the rate is utilized for mMIMO, whereas the upper bounds are used for the NOMA scheme. Notice that there is a significant difference between the rates achieved by the mMIMO and NOMA schemes for the cell-edge users. The reason that mMIMO outperforms NOMA at the cell-edge is that mMIMO provides the array gain to both users, while the NOMA scheme only gives it to the cell-center user.

Next, we compare the performance of the mMIMO and NOMA schemes in terms of sum rate. Consider the sum rate of a group under the NOMA scheme and assume for simplicity that indices $1,2$ represent the cell-center and the cell-edge users, respectively. Then, the sum rate for the two users is  
\begin{eqnarray}
R_{\text{sum}}^{\mathrm{NOMA}} = \tau\log_2\left(1 + \bar{M}p_1\beta_1 \right) + \tau\log_2 \left(1 + \frac{p_2\beta_2}{1 + p_1\beta_2}\right),
\end{eqnarray}  
where  $\bar{M} = M + 1 - K/2$. The maximization of this sum rate is investigated in \cite{choi2016power} in the two-user case.
Next, we generalize the two-user setup to $K$ users. Let $\vect{p} = [p_1,\ldots,p_K]^T$ denote the power vector. Then, the sum rate optimization problem can be stated as
      \begin{equation} \tag{P1}\label{pr:PR-1}
\begin{aligned}
& \underset{\vect{p}\succeq 0}{\text{maximize}}
& & \sum_{k = 1}^KR_k \\
& \text{subject to} 
& &  \sum_{k = 1}^{K} p_{k} \leq p_{\max}. 
\end{aligned}
\end{equation}

For the mMIMO scheme the optimization problem can be solved via the conventional water-filling algorithm described in \cite{telatar1999capacity}. 
For the NOMA scheme, the optimization problem has a more complicated structure, but we can prove the following.
\begin{lemma}\label{lem:NOMAlem1}
	For the NOMA scheme, the optimization problem \eqref{pr:PR-1} is maximized if and only if 
	\begin{equation}
	p_k = 0,\quad \forall k \in \mathcal{K}_e.
	\end{equation} 
\end{lemma}
\begin{IEEEproof}
	The proof is given in the journal version \cite{Senel2019a}. 
	\end{IEEEproof}
Lemma \ref{lem:NOMAlem1} proves that to obtain the maximum sum rate, the NOMA scheme does not allocate any power to the cell-edge users, which are effectively dropped from service. Notice that this result holds for any number of antennas and users. Hence, the NOMA scheme turns into a mMIMO scheme that only serves the cell-center users, which shows that NOMA is suboptimal. The power allocation for the cell-center users can be solved using the conventional water-filling algorithm.

	\begin{figure}[tb]
	\begin{center}
		\includegraphics[trim=0.5cm 0cm 0cm 0cm,clip=false, scale = .6]{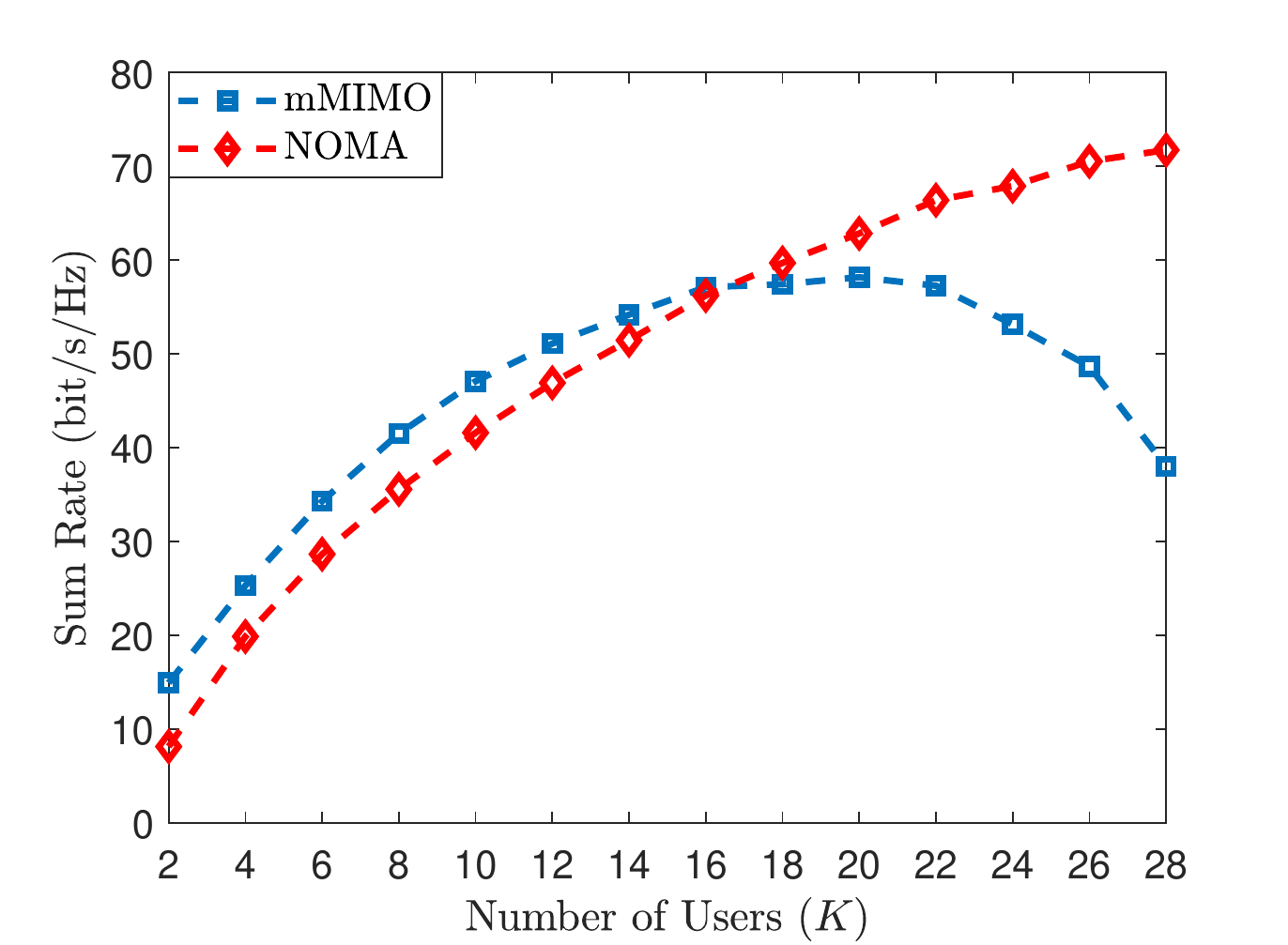}
		\caption{ The average sum rate obtained for the problem defined in \eqref{pr:PR-1} where the average is taken with respect to random user locations. The curves are obtained for various number of users under a setup with $M = 30$ BS antennas and coherence interval length $T = 100$ symbols. }
		\label{fig:Fig3} 
	\end{center}  \vspace{-5mm}
\end{figure}

In Fig.~\ref{fig:Fig3}, the average sum rates obtained by mMIMO and NOMA, with respect to the number of users, are depicted under a setup with $M = 30$ BS antennas. The cell-center and edge users are uniformly distributed  in certain parts of the cell such that the SNR of cell-center users are in the range 15--26\,dB whereas for cell-edge users it is from $-5\,$dB to $15\,$dB, when assigning full power to a user.
The curves represent the solutions to \eqref{pr:PR-1}. The figure shows that mMIMO provides a higher sum rate than NOMA for $K \leq 16$, while NOMA outperforms mMIMO scheme when $K > 16$. Hence, as $K/M \rightarrow 1$ NOMA becomes superior whereas when $M \gg K$, which is normally what is considered to be mMIMO in the literature, the mMIMO scheme outperforms the NOMA scheme. In other words, NOMA can outperform small-scale multi-user MIMO systems, but not a true mMIMO system.

	\begin{figure}[tb]
	\begin{center}
		\includegraphics[trim=0.5cm 0cm 0cm 0cm,clip=false, scale = .6]{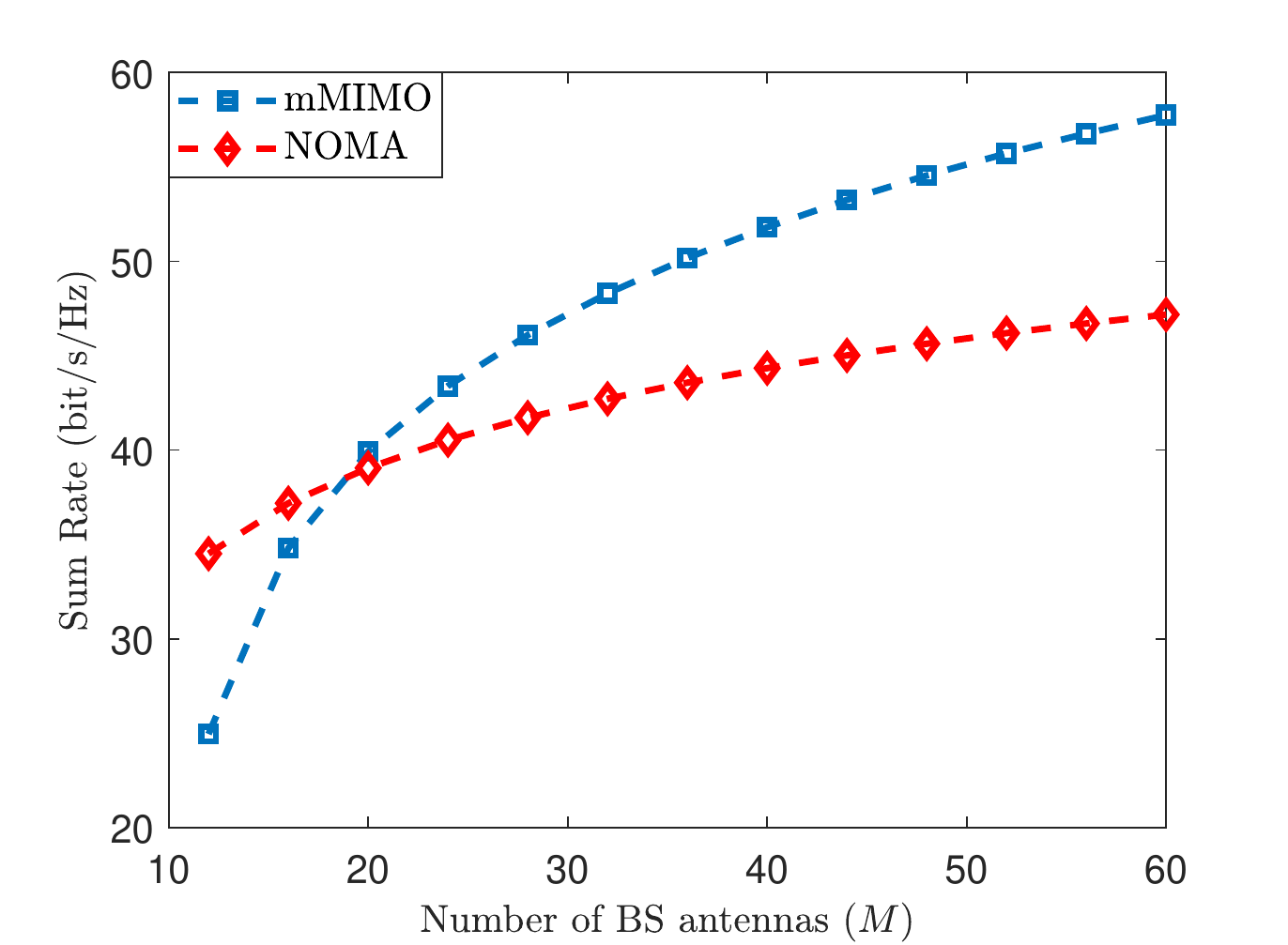}
		\caption{The curves represent the average sum rate obtained for the problem defined in \eqref{pr:PR-1} where the average is taken with respect to random user locations. The average sum rates are obtained for the mMIMO and NOMA schemes as a function of number of BS antennas for $K = 10$ users with a coherence interval length of $T = 100$ symbols.  }
		\label{fig:Fig4} 
	\end{center}  \vspace{-5mm}
\end{figure}
 
The performance of mMIMO and NOMA with respect to the number of BS antennas is shown in Fig.~\ref{fig:Fig4} for $K=10$. As expected, NOMA provides the highest performance  when $K/M \approx 1$, while when approaching the massive MIMO regime, mMIMO becomes significantly better than NOMA. Furthermore, even though both approaches benefit from a higher number of BS antennas, the difference between them increases with $M$. 

\subsection{Analytical Example: Two Users}

To analytically determine for which values of $M$ and $K$ that mMIMO and NOMA are preferable, we now  investigate \eqref{pr:PR-1} in the two-user case.
\begin{lemma}\label{lem:ZF_2user}
For the mMIMO scheme, \eqref{pr:PR-1} is maximized in the  two-user case by
	\begin{eqnarray}
	p_1 &=& \min\left(p_{\max},~\frac{\beta_1 - \beta_2 + p_{\max}\beta_1\beta_2\left(M - 2\right)}{2\beta_1\beta_2\left(M-2\right)}\right), \label{eq:ZFoptP1}\\
	p_2 &=& p_{\max} - p_1. \label{eq:ZFoptP2}
	\end{eqnarray}  	
\end{lemma}
\begin{IEEEproof}
	The proof is given in the journal version \cite{Senel2019a}. 
\end{IEEEproof}	

Hence, the maximum sum rate with mMIMO is 
\begin{equation} \label{eq:ZFmax_2user}
R_{\max}^{\mathrm{mMIMO}} = \tau \log_2\left(\frac{\left(\beta_1 + \beta_2 + p_{\max}\beta_1\beta_2\left(M -2\right)\right)^2}{4\beta_1\beta_2}\right),
\end{equation}
if $p_{\max} \geq \frac{\beta_1- \beta_2}{\beta_1\beta_2\left(M-2\right)}$. Otherwise, 
\begin{equation} \label{eq:ZFmax_2user2}
R_{\max}^{\mathrm{mMIMO}} = \tau \log_2\left( 1 + p_{\max}\beta_1\left(M - 2\right)\right).
\end{equation}

For the NOMA scheme, we have 
\begin{equation} \label{eq:NOMAmax_2user}
R_{\max}^{\mathrm{NOMA}} = \tau\log_2\left(1+ p_{\max}\beta_1 M\right).
\end{equation}
In the NOMA scheme, the sum rate scales with $M$ while with mMIMO it scales with $\left(M - 2\right)^2$ when both users are active. This suggest that when $M$ is small, NOMA may provide a better performance, however mMIMO will always outperform NOMA scheme when $M$ becomes sufficiently large. More precisely, $R_{\max}^{\mathrm{mMIMO}} \geq R_{\max}^{\mathrm{NOMA}}$, for $M \geq M^*$ where 
	\begin{equation} \label{eq:Mstar}
	M^* =  2 + \frac{\beta_1 - \beta_2}{p_{\max}\beta_1\beta_2} + \frac{2\sqrt{2}}{\sqrt{p_{\max}\beta_2}}.
    \end{equation}

\begin{figure}[tb]
	\begin{center}
		\includegraphics[trim=0.5cm 0cm 0cm 0cm,clip=false, scale = .6]{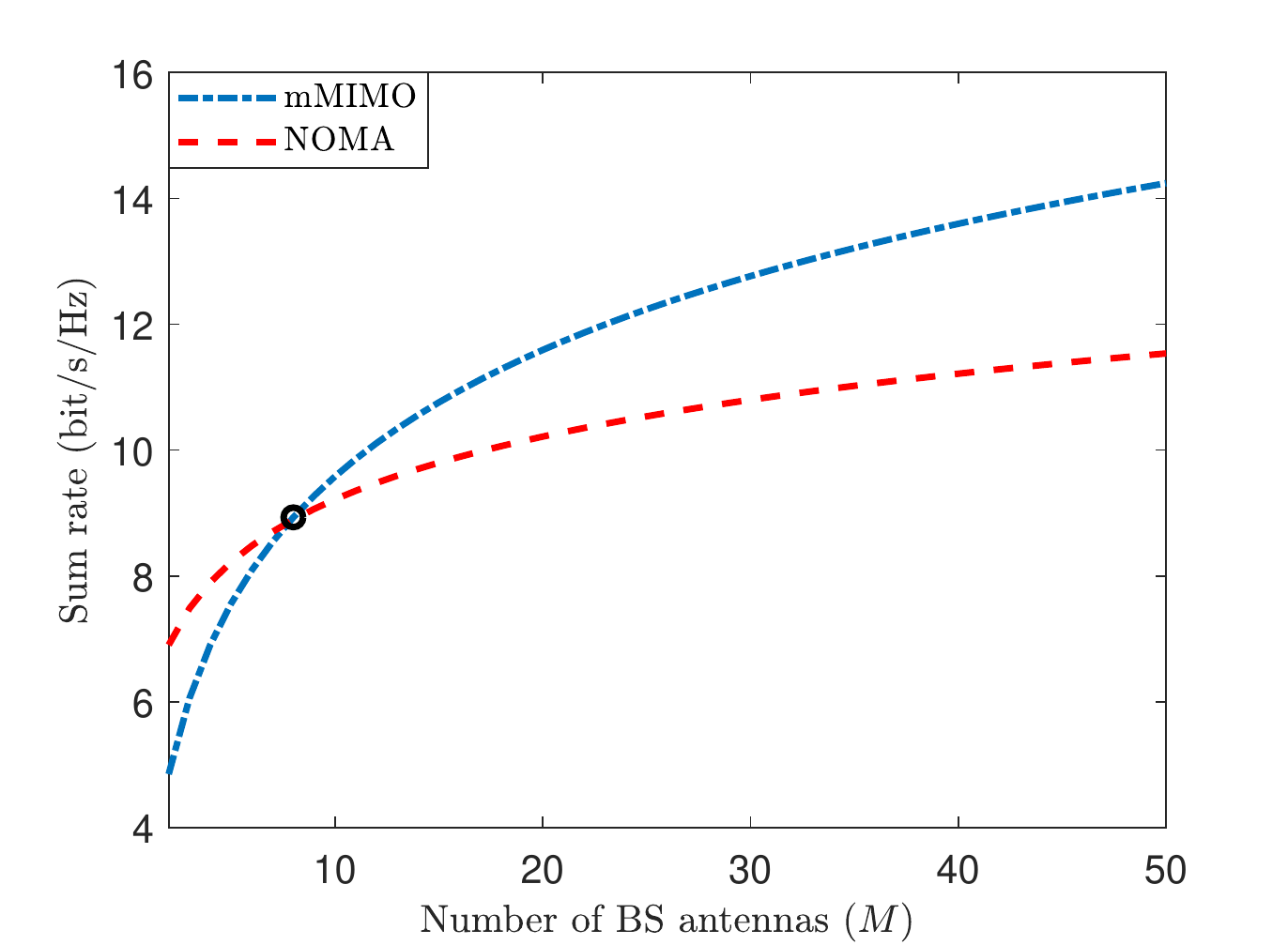}
		\caption{ Maximum sum rate as a function of number of BS antennas. Black circle shows $M^*$ defined in \eqref{eq:Mstar}. }
		\label{fig:Fig2} 
	\end{center}
\end{figure}

From \eqref{eq:Mstar}, we can see that NOMA is preferable when the difference between the large-scale fading coefficients is large. Second, a smaller $\beta_2$ also favors the NOMA scheme which is in alignment with the grouping of cell-edge and cell-center users. Finally, high transmit power favors the mMIMO scheme. Fig.~\ref{fig:Fig2} illustrates the sum rates obtained in the two-user case as well as $M^*$ given in \eqref{eq:Mstar}. We notice that mMIMO becomes superior for $M\geq 9$, which is much smaller than what is normally referred to as massive MIMO. In this particular example, we use the same SNRs as in Fig.~\ref{fig:Fig1}.

\section{Conclusion}

In this work, we have investigated the gains that NOMA can provide in mMIMO setups, which are characterized by having a large number of antennas compared to the number of users, and compared it with a standard ZF beamforming approach from the mMIMO literature. This comparison is  practically important since BSs equipped with 64 antennas have already been deployed in LTE networks and having massive antenna numbers will be the norm when the 5G deployments take off.

We proved analytically that when $M\gg K$, mMIMO achieves the highest average sum rate. However, in cases where $M \approx K$, the NOMA scheme can be preferable since ZF cannot be used to suppress all the interference. A closed-form expression was derived for the minimum number of BS antennas such that the mMIMO scheme outperforms NOMA, in the two user case, which shows that NOMA benefits from having a smaller large-scale fading coefficient for the cell-edge user and a large difference between the large-scale fading coefficients. Similar observations are made for line-of-sight channels in the journal version \cite{Senel2019a} of this paper, thus it is not limited to the i.i.d.~Rayleigh fading model used herein.

The intuition behind our result is that the favorable propagation (i.e., nearly orthogonal channel vectors) that appear in mMIMO systems makes NOMA inferior to conventional mMIMO schemes. In the journal version \cite{Senel2019a}, we show that there are cases where favorable propagation does not occur and then a hybrid between NOMA and conventional mMIMO can improve the performance, but the gains seem to be small.

\bibliographystyle{IEEEtran}
\bibliography{IEEEabrv,References}

\end{document}